\documentclass[10pt,conference]{IEEEtran}
\usepackage{array,epsfig,rotating,setspace,latexsym,amsmath,epsf,amssymb,bm,theorem}
\usepackage{cite,graphicx,multirow,color,caption,subcaption}
\usepackage{epstopdf}
\makeatletter
\def\ps@headings{%
	\def\@oddhead{\mbox{}\scriptsize\rightmark \hfil \thepage}%
	\def\@evenhead{\scriptsize\thepage \hfil \leftmark\mbox{}}%
	\def\@oddfoot{}%
	\def\@evenfoot{}}
\makeatother
\usepackage[top=0.75 in, bottom = 1 in, left= 0.625 in, right=0.64 in]{geometry}
\newtheorem{Theo}{Theorem}

\newtheorem{Def}{Definition}

\allowdisplaybreaks

\title{Covert Bits Through Queues\thanks{This work was supported by NSF Grants CNS 13-14733, CCF 14-22111, CCF 14-22129 and CNS 15-26608.}}
\date{}

\author{Pritam Mukherjee \qquad Sennur Ulukus\\
	\normalsize Department of Electrical and Computer Engineering\\
	\normalsize University of Maryland, College Park, MD 20742\\
	\normalsize {\it pritamm@umd.edu} \qquad {\it ulukus@umd.edu}}

\def\A{\mathbf{A}}
\def\D{\mathbf{D}}
\def\E{\mathbf{E}}

\allowdisplaybreaks
\begin{document}
	\IEEEoverridecommandlockouts
	\maketitle
	\begin{abstract}
			We consider covert communication using a queuing timing channel in the presence of a warden. The covert message is encoded using the inter-arrival times of the packets, and the legitimate receiver and the warden observe the inter-departure times of the packets from their respective queues. The transmitter and the legitimate receiver also share a secret key to facilitate covert communication. We propose achievable schemes that obtain non-zero covert rate for both exponential and general queues when a sufficiently high rate secret key is available. This is in contrast to other channel models such as the Gaussian channel or the discrete memoryless channel where only $\mathcal{O}(\sqrt{n})$ covert bits can be sent over $n$ channel uses, yielding a zero covert rate.  
	\end{abstract}
	
	\section{Introduction}
\begin{figure}[t]
		\centering
		\includegraphics[width=\linewidth]{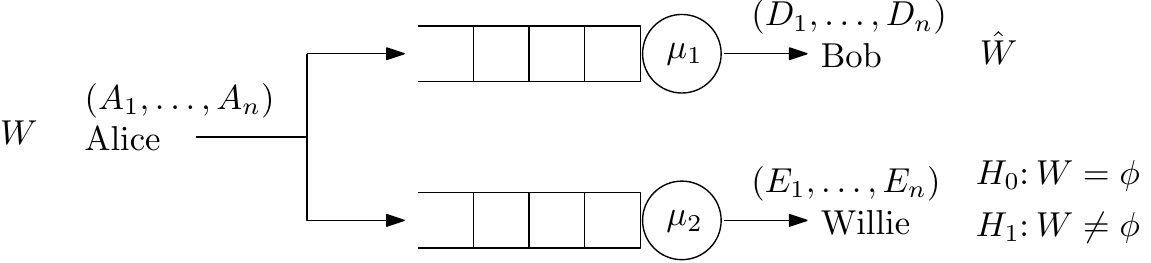}
		\caption{Covert communication over timing channels.}
		\label{fig:model}
		\vspace{-10 pt}
	\end{figure}

	We consider a covert communication system, where Alice wishes to send a message $W$ covertly to Bob using a timing channel in the presence of a warden Willie, as depicted in Fig.~\ref{fig:model}. To facilitate such covert information transfer, we allow Alice and Bob to have a shared secret key $K$. The channels to Bob and Willie are error-free bit pipes leading to buffers that are modeled as single-server queues with service rates of $\mu_1$ packets per second and $\mu_2$ packets per second, respectively. The packets arrive at both Bob's and Willie's queue simultaneously, and do not contain any covert information. When Alice does not have any covert messages to send, the arrival of the packets at the queues is modeled as a Poisson process with rate $\lambda$. In order to send the covert message $W$, Alice encodes $W$ in the inter-arrival times of the packets at the queues.  The warden Willie is a passive observer and tries to detect the presence of a covert message based on any \emph{unusual patterns} in the timings of the packets observed by him. The goal of this paper is the study strategies for reliable and covert information transmission from Alice to Bob in the presence of Willie.
	
	To that end, we first\footnote{Later, in Theorem 2, we consider queues with general service distributions.} assume that both queues have exponential service times; i.e., the service times of Bob's and Willie's queues are exponentially distributed with means $\frac{1}{\mu_1}$ and $\frac{1}{\mu_2}$, respectively. We exploit a result in \cite{btq96}, which characterizes the maximum rate of reliable information transmission, or the \emph{capacity}, using the timing channel with an exponential service time queue. This gives us an upper bound on the \emph{capacity} of the covert information transmission rate. Next, to ensure covertness, we want that the probability distribution of the departure times of the packets at Willie's queue remains almost the same, irrespective of whether a covert message is being sent or not. Thus, the distribution of the departure times at Willie's queue, induced by the designed codebook and the uniform choice of the covert message, closely \emph{approximates} the distribution of the departure times when the packet arrivals are modeled as a Poisson process with rate $\lambda$. This brings us to the setting of channel resolvability \cite{output_statistics}, where we want to approximate the output distribution induced by a process with the output distribution induced by a codebook. The results of \cite{btq96} and  \cite{output_statistics} together yield the proposed achievable schemes. 
	
	An interesting aspect of the result is the fact that a strictly positive rate is achievable in this case. This is in contrast to other covert channels studied in the literature, such as the covert Gaussian channel \cite{gaussian_covert} and the covert discrete memoryless channel \cite{bloch_covert,wornell_covert}, where only\footnote{$f(n)= \mathcal{O}(g(n)) \Leftrightarrow$ $\exists M,n_0$ s.t. $f(n)\leq Mg(n)$, $\forall n\geq n_0$.} $\mathcal{O}(\sqrt{n})$ bits can be sent in $n$ channel uses, i.e., the covert rate is zero. For each of these channels, there is an \emph{innocent} symbol which is transmitted in the absence of a covert message. In the Gaussian channel, for example, the default input symbol is zero when no communication is taking place. Analogously, for the discrete memoryless channel, it is a symbol $x_0$. In order to send information, one has to use at least one non-innocent symbol $x_1$; however, if too many\footnote{$f(n)= \omega(g(n)) \Leftrightarrow$ $\forall m>0, \exists n_0>0$ s.t. $f(n)>mg(n)$, $\forall n\geq n_0$.} (more than $ \omega(\sqrt{n})$) of the non-innocent symbol $x_1$ are sent in $n$ channel uses, the induced output distribution differs significantly from the default output distribution induced by the input of the innocent symbol only. In the Gaussian channel, for example, using too many non-zero symbols with non-vanishing power significantly increases the output power, and the warden can detect the communication. 
	
	On the queuing timing channel, packets naturally arrive at the queue with i.i.d.~exponential inter-arrival times when there is no communication using the timing channel. That is, the innocent state is a sequence of packets with i.i.d.~exponential inter-arrival times. On the other hand, when communication is taking place on the covert channel, the inter-arrival times belong to a codeword  drawn from a codebook. Thus, in order to detect whether communication is taking place, the warden tries to determine if the input to the timing channel, i.e., the inter-arrival times, belong to an i.i.d.~exponential process, or to a codeword drawn from a fixed codebook, using its output observations. Thus, to ensure covertness, we only need to ensure \emph{stealth} \cite{kramer_stealth}, which requires the output distribution induced by the codebook to approximate the default output distribution induced by an i.i.d.~input process closely.  Note that stealth does not imply covertness in Gaussian or discrete memoryless channels, since the warden can detect the transmission even if it cannot distinguish whether the received symbols belong to some codebook, or are outputs corresponding to i.i.d.~inputs by observing the output distribution. In the timing channel, however, stealth implies covertness, and as in discrete memoryless channels, the stealth constraint can be met with a strictly positive rate  \cite{kramer_stealth}. Hence, we achieve covertness on the timing channel with a non-zero rate.

	\emph{Related Work:} Timing channels have been widely investigated in the context of both communication and computer systems. In most cases, the timing channel is not the primary intended means of information transfer, and timing is not usually considered a data object. Thus, timing channels, by themselves, are considered \emph{covert} in most of the literature.
	
	 From a communication perspective, references \cite{moscowitz92,moscowitz96} provide achievable schemes for certain timing channel models and analyzes their performance using an information theoretic framework. Reference \cite{btq96} analyzes the limits of reliable information transmission using the timings of packet arrivals and departures, i.e., the capacity of the timing channel for a single-server queue. In contrast to usual discrete memoryless channels, the timing channel model used in \cite{btq96} has memory and is not stationary. Therefore, reference \cite{btq96} employs information spectrum methods \cite{han_verdu}  in order to analyze its capacity. The exact capacity is derived for queues with exponentially distributed service times, and upper and lower bounds are provided for general queues. While reference 
	\cite{btq96} deals with continuous time arrivals, reference \cite{badekar_azizoglu} analyzes discrete time queues and shows that the geometric service time distribution plays a role analogous to the exponential distribution for continuous time queues. Extensions to secure transmission of information using the timing channel are also available in the literature \cite{securebtq}. References \cite{kiyavash_coleman2008,kiyavash_coleman2009,sundaresan_verdu} study practical code designs that approach the capacity of the timing channel.   
	
	In the context of computer systems, a timing channel analysis can often represent a side channel attack from a malicious adversary. For example, in shared event schedulers, a malicious user process can infer information about the legitimate user process' arrival patterns, and compromise the privacy or security of the legitimate user. References \cite{parv_kiyavash2011,gong_kiyavash} quantify the information leakage in the context of such shared event schedulers. Mitigation of such timing channel attacks has also been studied in the literature, e.g., in references \cite{parv_kiyavash2012,gong_parv_kiyavash2012,parv_kiyavash_2013}. Reference \cite{hajek_giles} studies the problem of information leakage through the timing channel in the framework of a game between the covert transmitter-receiver pair and a jammer who tries to disrupt the covert communication while being subject to buffer or delay constraints. Other mitigation techniques such as predictive mitigation \cite{predictive_mitigation} and time-deterministic replay \cite{chen_haeberlen} have also been explored in the literature.         
	
	\section{System Model}
	We consider covert communication through two parallel single-server queues as shown in Fig.~\ref{fig:model}. The information in the packets is intended for both Bob and Willie. Alice also wants to send a covert message $W$ to Bob without allowing the warden Willie to detect it. The covert message is encoded in the arrival times of the packets entering both Bob's queue and Willie's queue simultaneously. In the absence of a covert message, the packet arrivals are modeled as a Poisson process with rate $\lambda$, i.e., the inter-arrival times of the packets are exponential with mean $\frac{1}{\lambda}$. Bob's queue has a service rate of $\mu_1$ packets/s, while Willie's queue has a service rate of $\mu_2$ packets/s.  We assume $\lambda \leq \min(\mu_1,\mu_2)$ to ensure stability of both queues.
	
	In order to send the covert message $W$, Alice encodes it in the inter-arrival times of the packets at the queues, i.e., the covert message is encoded as a vector of $n$ non-negative inter-arrival times $\textbf{A} = (A_1,\ldots,A_n)$, such that the $k$th packet enters both queues at time $\sum_{i=1}^k A_i$.  The intended receiver Bob and the warden Willie both observe the sequence of departure times of the packets at their respective queues. Therefore, the output of each channel is a vector of $n$ non-negative inter-departure times, which we denote by $\D = (D_1,\ldots D_n)$ and $\E = (E_1,\ldots,E_n)$ for Bob's and Willie's channels, respectively. The departure times of the $k$th packet from Bob's and Willie's queues are $\sum_{i=1}^{k}D_i$ and $\sum_{i=1}^{k}E_i$, respectively.
	
	The service times for the $k$th packet in Bob's and Willie's queues are denoted by $S_k$ and $T_k$, respectively. $S_k$ and $T_k$ are mutually independent of each other and of $\A$, $D^{k-1}$ and $E^{k-1}$. The \emph{idling time} for the $k$th packet is the time elapsed between the $(k-1)$th departure and the $k$th arrival; if the $k$th arrival occurs before the $(k-1)$th departure, the idling time is zero. Denoting the idling time for the $k$th packet in Bob's and Willie's queues by $U_k$ and $V_k$, respectively, we have,
	\begin{align}
	U_k =& \max\left(0, \sum_{i=1}^{k} A_i -\sum_{i=1}^{k-1} D_i\right)\\
	V_k =& \max\left(0, \sum_{i=1}^{k} A_i -\sum_{i=1}^{k-1} E_i\right)
	\end{align}  
	which are deterministic functions of $(A^{k}, D^{k-1})$ and $(A^{k}, E^{k-1})$, respectively. Also the inter-departure times can now be expressed as
	\begin{align}
	D_k =& U_k+S_k\\
	E_k =& V_k + T_k
	\end{align}
	We have the following Markov chains:
	\begin{align}
	&D_k \rightarrow U_k \rightarrow (A^k,D^{k-1})\\ &E_k\rightarrow V_k \rightarrow (A^k,E^{k-1})
	\end{align}
	Note that this model of the single-server queue has memory, a non-linear input-output relation and is non-stationary.
	
	The objective in the covert communication setting is to ensure that the warden Willie cannot detect the presence of a covert message with its observations of the departure times, while also allowing the intended receiver Bob to decode the covert message with probability of error approaching zero. To facilitate this, we also allow Alice and Bob to share a secret key $K$ which is not available to Willie. Formally, we have the following definition.

		\begin{figure}
			\centering
			\includegraphics[height=170 pt]{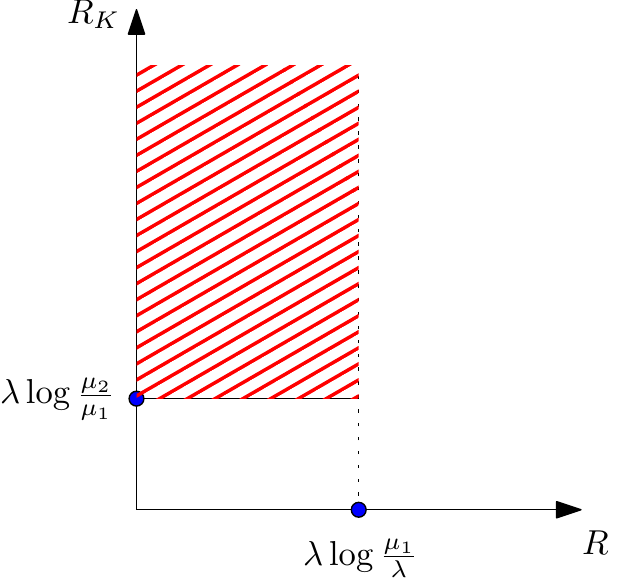}
			\caption{Achievable $(R,R_K)$ region when $\mu_2>\mu_1$.}
			\label{fig:strong_willie_region}
			\vspace{-5 pt}
		\end{figure}
	
	\begin{Def}
		An $(n,M_n,L_n,T_n,\epsilon_n, \delta_n)$ secret key assisted \emph{covert} timing code consists of the following:
		\begin{itemize}
			\item a message set $\mathcal{W}_n = \left\lbrace 1, \ldots, M_n\right\rbrace $ at Alice, from which a covert message $W$ is picked uniformly,
			\item a key set $\mathcal{K}_n =  \left\lbrace 1, \ldots, L_n\right\rbrace $ available at Alice and Bob, from which a secret key $K$ is picked uniformly,
			\item an encoding function (possibly stochastic) at Alice $\phi_n : \mathcal{W}_n \times \mathcal{K}_n \rightarrow \A $ that maps the message and the secret key into a codeword, which is a vector of $n$ non-negative inter-arrival times such that the $k$th arrival occurs at $\sum_{i=1}^{k}A_k$,
			\item a decoding function at Bob $\psi_n : \D \times \mathcal{K}_n \rightarrow \mathcal{W}_n$ that maps the observed codeword of inter-departure times $\D$ and the secret key $K$ to the decoded message $\hat{W}$, such that the probability of error 
			\begin{align}
			\mathbb{P}(W\neq \hat{W})\leq \epsilon_n
			\end{align}
			\item satisfies the covertness constraint at Willie,
			\begin{align}
			d(P_{\E},Q_0^n )\leq \delta_n
			\end{align}
			where $d(P,Q)$ denotes the variational distance $||P-Q||_1$, $P_{\E}$ denotes the distribution of $\E$ in the presence of covert message, $Q_0$ is the default distribution of inter-departure times when no covert message is present; in our case, if the arrival process has rate $\lambda < \min(\mu_1,\mu_2)$, $Q_0$ will also be a Poisson process with rate $\lambda$,
			\item the $n$th departure from Bob's queue occurs, on average, no later than $T_n$.
		\end{itemize}
	\end{Def}   
	
	A \emph{covert rate} $R$ is achievable with \emph{secret key rate} $R_K$, if there exists a $(n,M_n,L_n,T_n,\epsilon_n, \delta_n)$ \emph{covert timing code} with
	\begin{align}
	\liminf_{n\rightarrow \infty} \frac{\log M_n}{T_n} \geq& R\\
	\limsup_{n\rightarrow \infty} \frac{\log L_n}{n} \leq& R_K\\
	\limsup_{n\rightarrow \infty} \epsilon_n =& 0 \\
	\limsup_{n\rightarrow\infty} \delta_n =& 0
	\end{align} 
	
	A \emph{covert rate} $R$ is said to be achievable with \emph{secret key rate} $R_K$ at \emph{output rate} $\lambda$, if $R$ is achievable using a sequence of $(n,M_n,L_n,n\lambda,\epsilon_n, \delta_n)$ covert timing codes. The \emph{covert capacity at output rate} $\lambda$, denoted by $C(\lambda)$ is the supremum of all rates $R$ that are achievable with output rate $\lambda$.

		\begin{figure}
			\centering
			\includegraphics[height=170 pt]{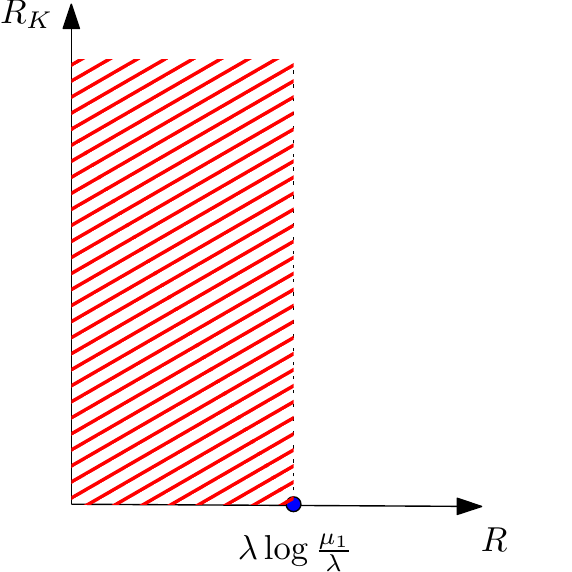}
			\caption{Achievable $(R,R_K)$ region when $\mu_2<\mu_1$.}
			\label{fig:weak_willie_region}
			\vspace{-5 pt}
		\end{figure}
	
	\section{Main Results}
    The main results of this paper are the following two theorems.
	\begin{Theo}\label{theo:main_theo}
		Assume both queues are $M/M/1$. Then, covert communication is possible with output rate $\lambda < \min(\mu_1,\mu_2)$ if $(R,R_K)$ lies in the following region:
		\begin{align}
		R\geq& 0\\
		R <& \lambda \log \frac{\mu_1}{\lambda}\\
		R_K >&  \max\left(0,\lambda\log \frac{\mu_2}{\mu_1}\right)
		\end{align}
	\end{Theo}
	
	Fig.~\ref{fig:strong_willie_region} shows the $(R,R_K)$ region when $\mu_2>\mu_1$, while Fig.~\ref{fig:weak_willie_region} shows the $(R,R_K)$ region when $\mu_2<\mu_1$. Note that when $\mu_2<\mu_1$, i.e., when the service rate of the Willie's queue is less than the service rate of Bob's queue, no positive rate secret key is required to achieve the maximum covert rate. Intuitively, Willie has a \emph{worse} timing channel in this case than Bob. On the other hand, when $\mu_2>\mu_1$, a secret key of sufficient rate is required for covertness, as shown in Fig.~\ref{fig:strong_willie_region}. 
	
	Further, note that even when $\mu_2>\mu_1$, given a sufficient rate of secret key to enable covert communication, Alice can communicate with Bob with the full capacity at output rate $\lambda$. In other words, there is no loss of rate due to the extra covertness constraints, as long as a secret key of sufficient rate is available to Alice and Bob. Intuitively, a secret key of sufficient length increases the size of the input codebook such that the output distribution induced by this codebook is close to that induced by an i.i.d.~input, thus, ensuring covertness. If a key of sufficient rate is not available, our achievable scheme cannot guarantee covertness. Thus, the minimum rate $R_{K}^{min}$ of the secret key, if required, can be considered the price of covertness, and is given by
	\begin{align}
	R_K^{min} = \max\left(0,\lambda\log\frac{\mu_2}{\mu_1} \right) 
	\end{align} 
	
	Also note the contrast of this result with the corresponding result for Gaussian channels \cite{gaussian_covert} or discrete memoryless channels \cite{bloch_covert}. In both these cases, no covert communication with positive rate is possible in general. The best achievable rate in $n$ channel uses is $O(\omega_n\sqrt{n})$ in each case, where 
	\begin{align}
	\lim_{n\rightarrow\infty} \omega_n =&0\\ \lim_{n\rightarrow\infty} n\omega_n =&\infty
	\end{align} 
	In each case, there is an \emph{innocent} symbol, the zero symbol for the Gaussian channel or some symbol $x_0$ for the discrete memoryless channel, which is transmitted in the absence of a covert message, and a corresponding \emph{default} output distribution. Hence, the default distribution at the input is the degenerate distribution with all the probability mass at the \emph{all-zero} (or, the all $x_0$) codeword. This degenerate distribution, by itself, cannot be used to transmit any information. In order to transmit information, at least one non-innocent symbol $x_1$ must be used. Using too many (more than $ \omega(\sqrt{n})$) such non-innocent symbols in $n$ channel uses, however, results in the induced output distribution to differ significantly from the default output distribution, and covertness is lost. In our case, however, we exploit the fact that packets naturally arrive at the queue even in the absence of a covert message, i.e., in our case, the \emph{default}, or \emph{innocent}, setting is a sequence of packets with i.i.d.~exponential inter-arrival times. Thus, we ensure \emph{stealth} \cite{kramer_stealth} in the timing channel. As in the discrete memoryless channel with a stealth constraint \cite{kramer_stealth}, we achieve a non-zero rate in the timing channel. Unlike discrete memoryless channels, however, stealth implies covertness in the timing channel.
	
	While the result in Theorem \ref{theo:main_theo} holds for timing channels with exponential service times, similar achievability results can be obtained for general service distributions as well. We have the following theorem.

		\begin{Theo}\label{theo:main_theo2}
			Assume that the queues are $M/G/1$, such that Bob's queue has a service distribution $P_B$ and Willie's queue has a service distribution $P_W$. Then, covert communication is possible with output rate $\lambda < \min(\mu_1,\mu_2)$ when $(R,R_K)$ lies in the following region:
			\begin{align}
			R\geq& 0\\
			R <& \lambda \log \frac{\mu_1}{\lambda}\\
			R+ R_K >& \max\left(0,\lambda \log \frac{\mu_2}{\mu_1} + \lambda D(P_W||e_{\mu_2}) \right)
			\end{align}
			where $e_{\mu_2}$ denotes the exponential distribution with mean $\frac{1}{\mu_2}$.
		\end{Theo}
		
		Note that the rates in the theorem above are sufficient conditions to ensure covert communications; they are not necessary. As in the case of the exponential queue, no secret key is required if $\mu_1$ is sufficiently larger than $\mu_2$: when $\log \frac{\mu_1}{\mu_2} > D(P_W||e_{\mu_2})$.
	
	\section{Proofs of Theorems \ref{theo:main_theo} and \ref{theo:main_theo2}}
		
	\subsection{Proof of Theorem \ref{theo:main_theo}} An achievable scheme for covert communication is as follows:
	
	\noindent \textbf{Encoding:} 	First, fix any $(R,R_K)$ in the achievable region stated in Theorem \ref{theo:main_theo}. Introduce a dummy message $\tilde{W} \in \tilde{\mathcal{W}} = \left\lbrace 1,\ldots, 2^{nR_0}\right\rbrace $, where $R_0$ is chosen such that 
	\begin{align}
	 R+R_0 \leq& \lambda\log\frac{\mu_1}{\lambda}\\
	 R+R_0+R_K \geq& \lambda\log\frac{\mu_2}{\lambda}
	\end{align}
	Note that this is possible as long as $(R,R_K)$ lies in the region specified in Theorem \ref{theo:main_theo}. Bob will try to decode $(W,\tilde{W})$. Then:	
	\begin{itemize}
		\item The transmitter generates a random codebook with $2^{n(R+R_0+ R_K)}$ $n$-length codewords, with each code symbol being drawn in an i.i.d.~fashion from an exponential distribution with mean $\frac{1}{\lambda}$.  The codewords are indexed as $\mathbf{a}(w,\tilde{w},k)$, $w\in \left\lbrace1,\ldots, 2^{nR} \right\rbrace $, $\tilde{w}\in \left\lbrace1,\ldots, 2^{nR_0} \right\rbrace $ and  $k\in \left\lbrace 1,\ldots, 2^{nR_K}\right\rbrace $.
		\item To send a message $w \in \left\lbrace1,\ldots, 2^{nR} \right\rbrace $, when the realization of the secret key is $k\in \left\lbrace 1,\ldots, 2^{nR_K}\right\rbrace $, the transmitter chooses $\tilde{w}$ uniformly from $\left\lbrace 1,\ldots, 2^{nR_0}\right\rbrace $, and sends $\mathbf{a}(w,\tilde{w},k)$, i.e., the inter-arrival time of the $k$th packet is the $k$th code symbol $a_k(w,\tilde{w},k)$. 
	\end{itemize}
	
	\noindent\textbf{Decodability:} Bob tries to decode $(w,\tilde{w})$. Since the secret key is available at Bob, decoding with vanishing probability of error is possible as long as 
	\begin{align}
	R+R_0 \leq \mbox{p-}\liminf_{n\rightarrow \infty} \frac{1}{n} i_{\A;\D}(\A;\D)
	\end{align}
	where 
	\begin{align}
	 i_{X;Y}(X;Y) = \frac{p_{XY}(X,Y)}{p_X(X)p_Y(Y)}
	\end{align} 
	is the \emph{mutual information density} random variable, and 
	\begin{align}
	&\mbox{p-}\liminf_{n\rightarrow \infty} \frac{1}{n} i_{X^n;Y^n}(X^n;Y^n) \nonumber\\&= \sup\left\lbrace r: \lim_{n\rightarrow\infty} \mathbb{P}\left[\frac{1}{n}i_{X^n;Y^n}(X^n;Y^n) < r\right] =0  \right\rbrace 
	\end{align}	
	It is shown in \cite{btq96} that in our case when $\lambda<\mu_1$
	\begin{align}
	\mbox{p-}\liminf_{n\rightarrow \infty} \frac{1}{n} i_{\A;\D}(\A;\D) = \lambda\log\frac{\mu_1}{\lambda}
	\end{align}
	Therefore, the decodability constraint is satisfied.
	
	\noindent \textbf{Covertness:} We note that the secret key is not available at Willie. However, we want to approximate the i.i.d.~random Poisson process with rate $\lambda$ by the output induced by our codebook in order to ensure covertness. From the result of \cite{output_statistics}, this is possible if
	\begin{align}
	R+R_0+ R_K \geq \mbox{p-}\limsup_{n\rightarrow \infty} \frac{1}{n} i_{\A;\E}(\A;\E)
	\end{align}	
	Again, it is shown in \cite{btq96} that in our case when $\lambda<\mu_2$
	\begin{align}
	\mbox{p-}\limsup_{n\rightarrow \infty} \frac{1}{n} i_{\A;\E}(\A;\E) = \lambda\log\frac{\mu_2}{\lambda}
	\end{align}
	Therefore, the covertness constraint is also satisfied. This completes the proof of Theorem \ref{theo:main_theo}.
	
	\subsection{Proof of Theorem \ref{theo:main_theo2}}
	The achievable scheme for the general queue is similar to the scheme for the exponential queue. The encoding involves a random codebook with $2^{n(R+R_0+R_K)}$ $n$-length codewords, with each code symbol being drawn in an i.i.d.~fashion from the exponential distribution with mean $\frac{1}{\lambda}$. To send a message $w \in \left\lbrace1,\ldots, 2^{nR} \right\rbrace $, when the realization of the secret key is $k\in \left\lbrace 1,\ldots, 2^{nR_K}\right\rbrace $, the transmitter chooses a dummy message $\tilde{w}$ uniformly from $\left\lbrace1,\ldots, 2^{nR_0} \right\rbrace $ and sends $\mathbf{a}(w,\tilde{w},k)$, i.e., the inter-arrival time of the $k$th packet is the $k$th code symbol $a_k(w,\tilde{w},k)$. 
	
	\noindent\textbf{Decodability:} As in the $M/M/1$ case, since the secret key is available at Bob, decoding $(w,\tilde{w})$ with vanishing probability of error is possible as long as
		\begin{align}
		R+R_0 \leq \mbox{p-}\liminf_{n\rightarrow \infty} \frac{1}{n} i_{\A;\D}(\A;\D)
		\end{align}
	It is known from \cite{btq96} that capacity of the timing channel with exponential service time is the least among all service distributions. Hence, decoding is guaranteed when 
	\begin{align}
	R+R_0 \leq \lambda\log \frac{\mu_1}{\lambda}\label{eq:decode}
	\end{align}
	\noindent\textbf{Covertness:} As in the $M/M/1$ case, covert communication is possible as long as 
		\begin{align}
		R+R_0+R_K \geq \mbox{p-}\limsup_{n\rightarrow \infty} \frac{1}{n} i_{\A;\E}(\A;\E)
		\end{align}
	Again from \cite{btq96}, it is known that for the $M/G/1$ channel, the capacity of the timing channel is bounded by  $\lambda \log \frac{\mu_2}{\lambda} + \lambda D(P_W||e_{\mu_2})$. Therefore, covertness is guaranteed as long as 
	\begin{align}
	R+R_0+ R_K \geq 	\lambda \log \frac{\mu_2}{\lambda} + \lambda D(P_W||e_{\mu_2})\label{eq:covert}
	\end{align} 
	
	Eliminating $R_0$ from the decodability condition in \eqref{eq:decode} and the covertness condition in \eqref{eq:covert} yields the rate constraints in Theorem \ref{theo:main_theo2}. This completes the proof of Theorem \ref{theo:main_theo2}.

	\section{Conclusions}
	We introduced the notion of covert communication using a queuing timing channel in the presence of a warden. The covert message is encoded using the inter-arrival times of the packets and the legitimate receiver and the warden observe the inter-departure times of the packets from their respective queues. The transmitter and the legitimate receiver also share a secret key to facilitate covert communication. We proposed achievable schemes that obtain a non-zero covert rate for both $M/M/1$ and $M/G/1$ queues. This is in contrast to other channel models such as the Gaussian channel or the discrete memoryless channel where only $\mathcal{O}(\sqrt{n})$ covert bits can be sent over $n$ channel uses, and therefore, the achievable covert rate is zero. We exploit the fact that \emph{stealth} implies \emph{covertness} in the timing channel.

	\bibliographystyle{unsrt}
	\bibliography{references}	
\end{document}